\documentclass[preprint,final,1p,article]{elsarticle}
\usepackage{lineno}
\modulolinenumbers[2]
\journal{Nuclear Physics A}
\usepackage{amssymb}
\usepackage{graphicx}
\usepackage{color}
\usepackage{amsthm}
\usepackage{amsmath}
\usepackage{amssymb}
\usepackage{multirow}
\usepackage[colorlinks,citecolor=blue,linkcolor=red,anchorcolor=blue,filecolor=blue,urlcolor=blue]{hyperref}

\journal{Nuclear Physics A}

\begin{document}

\begin{frontmatter}



\title{Cluster production in $^{129,124}$Xe + $^{112,124}$Sn collisions between 65–150 MeV/nucleon}


\author[1]{T. G{\'e}nard} 
\author[1]{A. Chbihi}
\ead{abdelouahad.chbihi@ganil.fr}

\author[1]{C. Ciampi}
\author[2]{D. Durand}
\author[1]{Q. Fable}
\author[3]{A.~Le~F{\`e}vre}

 \author[4,5]{S.~Mallik}

\address[1]{GANIL, CEA/DRF-CNRS/IN2P3, Bvd. Henri Becquerel, F-14076, Caen CEDEX, France}
                     
\address[2]{Normandie Univ, ENSICAEN, UNICAEN, CNRS/IN2P3, LPC Caen, France}

\address[3]{GSI Helmholtzzentrum fur Schwerionenforschung GmbH, D-64291, Darmstadt, Germany}

\address[4]{Physics Group Variable Energy Cyclotron Centre, 1-AF Bidhan Nagar, Kolkata, 700064 India}

\address[5]{Homi Bhabha National Institute Training School Complex Anushakti Nagar, Mumbai, 400085 India}

\begin{abstract}
Characterization of the participant-zone (PZ) in the $^{129,124}$Xe + $^{112,124}$Sn reaction at the energy range 65-150 MeV/nucleon reveals copious cluster production. A detailed study of the chemical composition as a function of the impact parameter shows that heavier clusters ($^6$He, $^{6-8}$Li, $^{7-10}$Be...) are most likely produced for central collisions. A hierarchy of the cluster production with the neutron-richness of the total system is observed, suggesting a full mixing of the projectile and target in the PZ. An estimate of the maximum density in central collisions has been deduced from the kinetic energy of the emitted fragments, reaching almost 2 time the normal density (2$\rho_0$) at 150 MeV/nucleon.
\end{abstract}







\end{frontmatter}


\section{Introduction}
\label{sec_intro}

Heavy-ion collisions are a unique tool in the laboratory for studying the nuclear equation of state (NEOS) of nuclear matter. In particular, the study of clusters is a key ingredient in better understanding and constraining the NEOS \cite{lattimer2016,ditoro2010,tsang2012,lattimer2013,huth2022, horowitz2014}. During the collision, the system is expected to undergo a compression followed by expansion, while a copious number of light and intermediate clusters are emitted  \cite{ono2019}. Nevertheless, it is still unclear whether cluster formation is favored during the compression or expansion phase  \cite{ono2019,cozma2017}. 
To address this issue, several transport models based on various assumptions and approximations attempt to predict cluster formation\cite{cozma2017,ono2020}, while there is still a lack of theoretical framework for studying nuclear dense matter. A better understanding of cluster production is also crucial when modeling compact astrophysical objects \cite{typel2015}. The present work will focus on the systematic characterization of clusters formed during Xe+Sn collisions, at incident energies between 65 and 150 MeV/nucleon. These data were collected at GSI and GANIL, using the INDRA multidetector array\cite{pouthas1995}. They are compared with a semi-classical model, ELIE  \cite{durand2008} and also BUU  \cite{buu1933,mallik2015}.
 
\section{Event selection and impact parameter estimation}
We selected the participant zone (PZ) by requiring a cut around 90$^{0}$ in the reaction center of mass (c.m.) frame (70$<$$\Theta_{c.m.}$$<$110). This cut has been optimized using the ELIE event generator \cite{durand2008}, in order to ensure that at least 90$\%$ of all produced particles in the selected area are participant products.
The quantity of charged products selected in PZ is expected to be linked to the centrality of the collision. In peripheral collisions, the overlap between projectile and target is small produces fewer particles, whereas in the central collisions, the overlap is very large and produces a large quantity of particles. Consequently, the reduced impact parameter, $b_{red}$,  was estimated using the sample of charged products selected in the PZ as
$b_{red} = \frac{\sum(Z_{PZ})}{Z_{PZ}^{tot}}$, where $Z_{PZ}$ is the charge of given particle of PZ and $Z_{PZ}^{tot}$ is the total charge of PZ. The relation between the later quantity and the reduced $b_{red}$ is established using the ELIE model. 
\begin{figure}
    \includegraphics[width=0.6\textwidth] {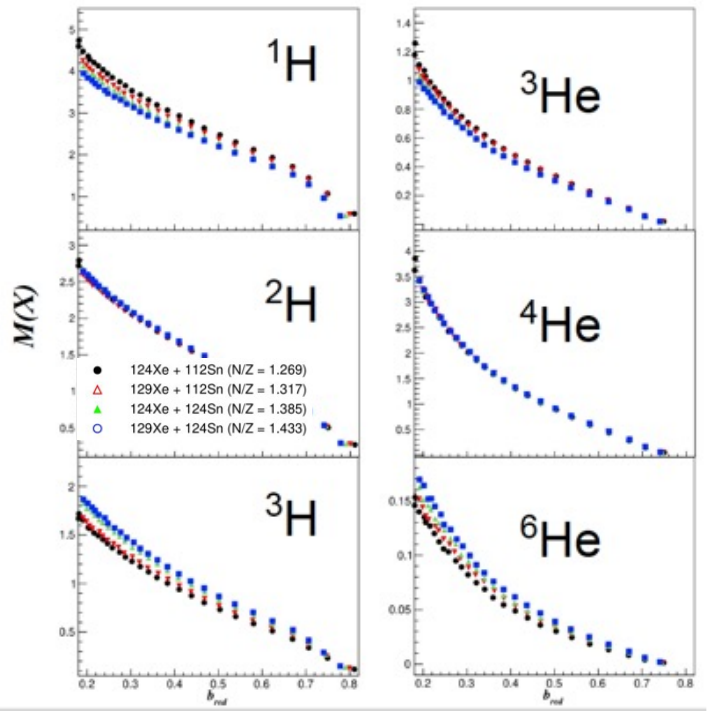}    
    \caption{ (color online) LCP multiplicities in the PZ vs $b_{red}$ obtained in the $^{129,124}$Xe + $^{112,124}$Sn systems at 100 MeV/nucleon.}
    \label{fig:multhhe}
\end{figure}
\section{Light charged particle (LCP) multiplicities}
\begin{figure}
\includegraphics[width=0.6\textwidth]{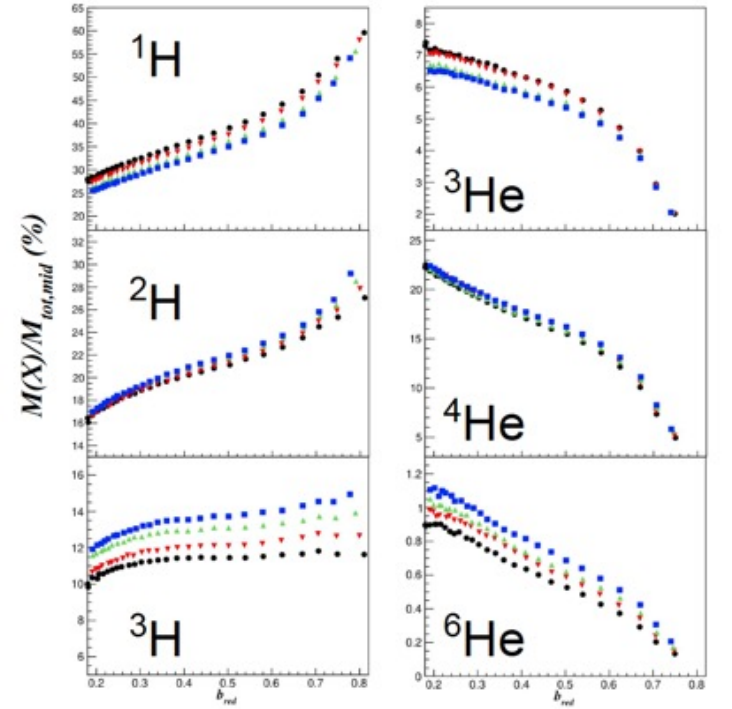}
    \caption{(color online) Normalized LCP multiplicities in the PZ vs b$_{red}$ obtained in the $^{129,124}$Xe + $^{112,124}$Sn systems at 100 MeV/nucleon.}
    \label{fig:normulthhe}
\end{figure}
Figures \ref{fig:multhhe} and \ref{fig:normulthhe} show the experimental multiplicities and normalized multiplicities of light charged particles (LCPs) emitted by the PZ  as a function of the reduced impact parameter for the four systems, $^{129,124}$Xe + $^{112,124}$Sn at 100 MeV/nucleon. As expected, protons and $\alpha$ particles are the most produced in this energy range, while $^{3}$He and $^{6}$He nuclei are less produced.  

The superposition of the multiplicity distributions of the four systems with different N/Z shows similar behavior. In addition, a dependence of the neutron hierarchy on the N/Z of the total system is observed. Proton and $^{3}$He neutron-poor particles are mainly emitted by neutron-poor systems, while $^{3}$H and $^{6}$He neutron-rich clusters are emitted by neutron-rich systems. For particles with the same number of neutrons and protons ($^{2}$H and $^{4}$He), there is little dependence on the four systems and all b$_{red}$. This neutron hierarchy suggests a full mixing of the projectile and target in the probed impact parameter domain. 

For each bin of the reduced impact parameter (b$_{red}$), the particle multiplicities were normalized to the total PZ multiplicity in that given bin. This normalization allows us not to depend  on the total mass of the four systems. The corresponding distributions are presented in Figure \ref{fig:normulthhe}. The evolution of the proportion of the total multiplicity of these particles in the participant zone depends on the nature of the particle. Hydrogen isotopes exhibit an increase with impact parameter, while helium isotopes show a negative slope (decrease with impact parameter). The transition between these two regimes appears to be mass dependent, with tritons serving as the transition point, as its distribution is almost  independent of impact parameter. Steeper decreases with the increase of b$_{red}$ are observed for Li and Be isotopes (not shown here). This effect suggests that, for central collisions (low b$_{red}$), heavy clusters (A>3) are preferentially produced to the detriment of light clusters in the participant-zone. To quantify this effect we have compared the ratio of the proportion of different particles produced at  b$_{red}$ = 0.2 and  0.6 , ($ \frac{ b_{red}=0.2}{ b_{red}=0.6}$). It equals to 0.7 for protons, 3.45 for $^{8}$Li and 14.0 for $^{10}$Be. The same phenomena was observed at higher incident energies by \cite{reisdorf2010}  where the increased stopping indicates increased compression. The increasing constrained transverse rapidity variances are interpreted as increasing radial flow subsequently developed in the expansion phase, which is coupled with increased cooling (‘droplet formation').

These study have been extended to other incident energies from 65 to 150 MeV/nucleon. Similar observations have been made. To conclude, heavy clusters are produced in central collisions in this energy range, suggesting that they are produced in rather dense nuclear matter.

\section{Dynamical characterization of the clusters in the participant zone}

The second part of this analysis focuses on the dynamical characterization of clusters in the participant zone.  
 \begin{figure}
   \centering
    \includegraphics[width=1. \textwidth]{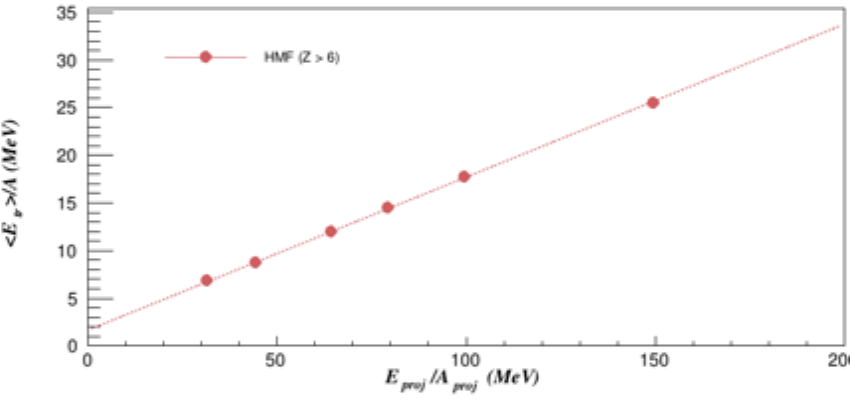}
    \caption{ Transverse energy as a function of incident energy.} 
    \label{fig:transverse}
\end{figure}   
As shown in Figure \ref{fig:transverse}, there is a strong linear correlation between the average transverse energy (E$_{tr}$) and the incident energy of the fragments emitted by the participant zone. The Figure only shows the case of Z = 6, the other intermediate mass fragments have same behavior. The origin of this correlation could be due to a collective motion of the fragments resulting from compression followed by expansion. We tested this scenario with the event generator ELIE \cite{durand2008} and the prescription of ref. \cite{daniel1979} utilizing the EOS and the compression energy with the following relationships:
$e_{tr} = e_{tr}(\rho_{0}) + e_{comp}(\rho)  $, where $ e_{comp} = K_{\infty}/18(\frac{\rho-\rho_{0}}{\rho_{0}})^{2} $. The parameterization of the maximum density the system can sustain is given by:
\begin{equation}
\frac{\rho}{\rho_{0}} = 1+ \sqrt{18\frac{(e_{tr}-e_{tr,th})}{K_{\infty}}}\left( \frac{A(b_{red})}{A_{0}}\right)^{\alpha}
\end{equation}
where $ e_{tr,th} $ is the threshold energy parameter (two values were used 30 and 45 MeV/nucleon),  $A(b_{red})$ is the size of the PZ at given reduced impact parameter, $ A_{0}$ is total size fixed at 240 and the $\alpha$ parameter fixed at 1/3.
We find that the maximum densities calculated largely exceed the saturation density $\rho_0$, up to 2 $\rho_0$, for an incident energy of 150 MeV/nucleon. These results are compatible with those of published transport models  BNV\cite{lefevre1997}, pBUU\cite{stone2017}, TDHF\cite{stone2017}. A calculation of the maximum density, performed with BUU@VECC-McGill\cite{mallik2020} reproduces very well the density extracted from Eq. 1.

\section{Conclusion}
The neutron-richness hierarchy on the total system indicates that the full mixing of the projectile and target in the participant zone is reached. Heavier clusters are most likely produced for central collisions, where the matter is expected to be compressed. A rough estimation of the maximal density reached, based on its relation with the experimental transverse kinetic energy is about 2$\rho_{0}$ at 150 MeV/nucleon. Our result is compatible with published transport models calculations and can provide important constraints.
 
\title{Acknowledgments}
The authors acknowledge the staff of the GSI for providing high quality
$^{124,129}$Xe beams adapted to the requirements of the INDRA multidetector.
This work was supported by the French-German Collaboration
Agreement 03-45 between IN2P3 - DSM/CEA and GSI.
 
%




\end{document}